\documentclass[aps,prl,twocolumn,superscriptaddress]{revtex4}
\usepackage{graphicx}
\usepackage{amssymb}
\usepackage{amsmath}
\usepackage{graphicx}
\usepackage{epsfig}
\usepackage{color}

\begin{document}

\title{Universal Scaling Law of Quasiparticle Nernst Effect in Cuprates:\\
A Unified Schematic Analysis for Transverse Transport}
\author{Yi-feng Yang}
\email[]{yifeng@iphy.ac.cn}
\affiliation{Beijing National Laboratory for Condensed Matter Physics and Institute of Physics,
Chinese Academy of Sciences, Beijing 100190, China}
\affiliation{School of Physical Sciences, University of Chinese Academy of Sciences,
Beijing 100190, China}
\affiliation{Songshan Lake Materials Laboratory, Dongguan, Guangdong 523808, China}
\date{\today }

\begin{abstract}
We discover a universal scaling law for the quasiparticle Nernst coefficient in underdoped cuprates, whose magnitude decreases exponentially with increasing temperature as confirmed in YBa$_2$Cu$_3$O$_y$. We attribute it to the basic mathematical structure of the conductivity formula with a narrow effective bandwidth of nonzero Berry curvatures associated with the pseudogap. A unified scheme is then developed to analyze transverse transport in hole-doped cuprates and clarify the puzzling disparity in determining the pseudogap temperature from different measurements. Our proposal opens the avenue for exploring potential scaling laws in the intermediate temperature region and may have broad applications in strongly correlated or narrow band systems.
\end{abstract}

\maketitle
The Nernst effect measures the transverse electric field generated by a longitudinal thermal gradient under a perpendicular magnetic field \cite{Nernst1886,Behnia2009}. It is typically  small in simple metals due to the so-called Sondheimer cancellation \cite{Sondheimer1948}. Investigations of the Nernst effect in correlated materials was greatly stimulated by the study of cuprate superconductors \cite{Xu2000,Wang2003}, where large Nernst signals have been reported in the pseudogap phase and attributed to vortices or votexlike excitations. While most studies have focused on the superconducting contribution \cite{Behnia2016RPP}, a relatively small quasiparticle term has later been identified and used to reveal nematicity of the pseudogap \cite{Daou2010Nature}. In spite of its importance, a basic understanding is still lacking for the quasiparticle Nernst effect. In particular, there is still no mathematical expression to describe its temperature evolution, let alone the underlying cause.

We report here the discovery of a universal scaling law for the quasiparticle contribution to the Nernst effect in hole-doped cuprates, and propose a simple scaling analysis based on the generic mathematical structure of the conductivity formula. The scaling is attributed to some potential topological properties of charge carriers in a finite energy window associated with the pseudogap. Comparison with experiments confirms our analysis and suggests a unified scheme to account for the transverse transport properties such as the Hall coefficient and the Nernst coefficient. We clarify the puzzling disparity in determining the pseudogap temperature from different experimental probes. Our work highlights the peculiar scaling properties of intermediate temperature physics and may have far-reaching implications in understanding transverse transport in correlated systems.

\begin{figure}[t]
\centering\includegraphics[width=0.45\textwidth]{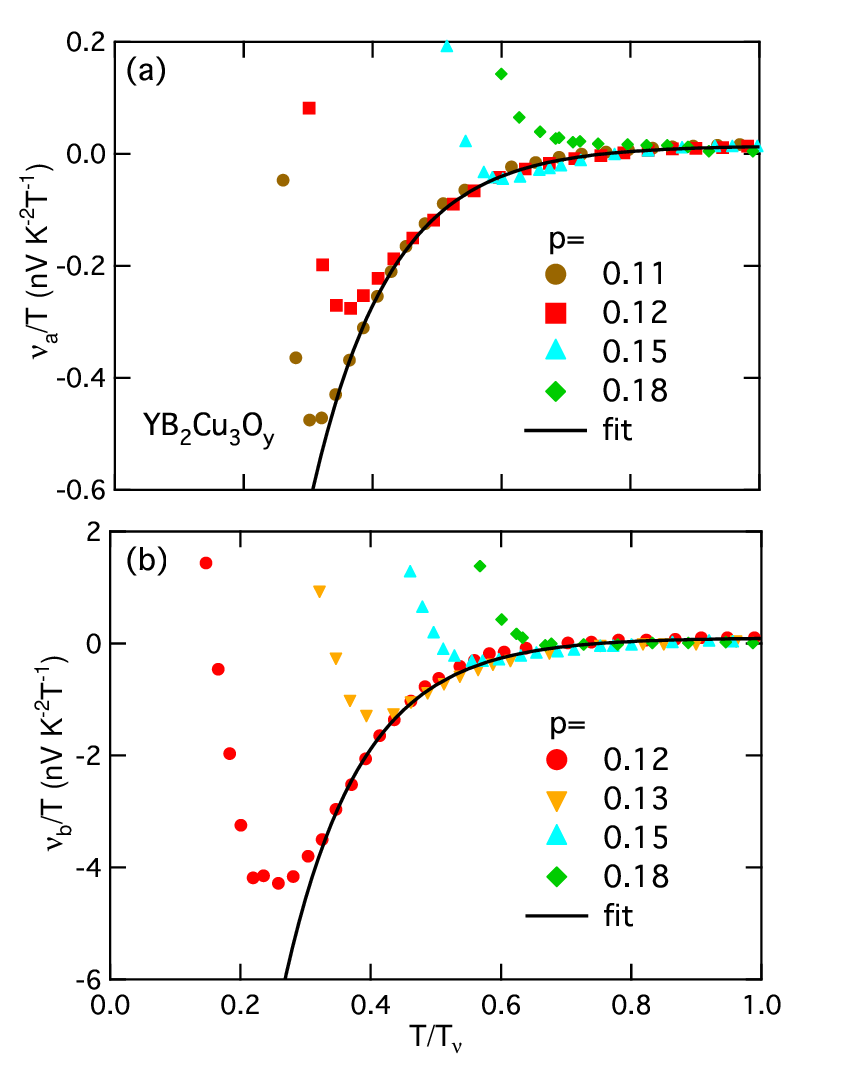}
\caption {Nernst coefficient along two perpendicular in-plane directions in YBa$_2$Cu$_3$O$_y$ reproduced from experiment \cite{Daou2010Nature}. $p$ denotes the effective hole doping. $T_\nu$ is the identified onset temperature of the Nernst coefficient assigned to the pseudogap. The solid lines are the exponential function, $e^{-T/T_0}$, where $T_0$ is roughly $T_\nu/8$.}
\label{fig1}
\end{figure}

We start with the Nernst measurement in YBa$_2$Cu$_3$O$_y$ (YBCO) by Taillefer's group \cite{Daou2010Nature}. The data are reproduced in Fig. \ref{fig1} and show an abrupt upturn at low temperatures. The large positive contribution is attributed to superconducting fluctuations, while the smaller negative contribution at higher temperatures comes from the quasiparticles. We have therefore two components, $\nu/T=\nu^{\text{sc}}/T+\nu^{\text{qp}}/T$. The presented data have been scaled with respect to some onset temperature $T_\nu$ assigned to the pseudogap \cite{Daou2010Nature}. Comparison of the data in Figs. \ref{fig1}(a) and \ref{fig1}(b) reveals a large in-plane anisotropy of the Nernst coefficient, suggesting possible nematicity in the pseudogap phase. We will not discuss this issue here, but focus on their overall temperature dependency. 

Our discovery is that the quasiparticle Nernst coefficient exhibits an exponential scaling at intermediate temperatures:
\begin{equation}
\nu^{\text{qp}}/T=A + B e^{-T/T_0},
\label{RHscaling}
\end{equation}
where $T_0$ is a characteristic temperature and $A$, $B$ are free parameters. Despite of the large anisotropy, all data can be well fitted using the exponential function (solid line) that covers exactly the region where quasiparticle contribution dominates, and yields similar temperature scale $T_0\approx T_\nu/8$, with $T_\nu$ being about 220 K for $p=0.12$ ($y=6.67$) assigned as the onset temperature of the pseudogap. Such a scaling can also be easily verified in other cuprate compounds.

To understand its origin, we study the general formula for transverse transport \cite{Yang2020PRL,Yang2023arXiv}:
\begin{equation}
\frac{\sigma^\alpha_{xy}}{H}=\frac{4\pi^2}{3}\int d\omega\frac{-\partial f(\omega)}{\partial \omega}\left(\frac{\omega}{T}\right)^\alpha \mathcal{B}(\omega,T),
\end{equation}
where $f(\omega)$ is the Fermi-Dirac distribution function and $\mathcal{B}(\omega,T)$ is determined by specific transport mechanism. $\alpha=0$, 1, 2 stands for the Hall conductivity ($\sigma_{xy}$), the off-diagonal Peltier coefficient ($\alpha_{xy}$), and the thermal Hall conductivity ($\kappa_{xy}$). The linear-in-field approximation is valid for the quasiparticle contribution since $\nu/T$ is almost unchanged for field up to 15 T in the scaling region \cite{Daou2010Nature}. The term $-\partial f(\omega)/\partial\omega$ accounts for the effect of thermal broadening and will be shown to play an essential role in causing the exponential scaling law.

The above formula covers a wide variety of possibilities where $\mathcal{B}(\omega,T)$ may have different origins and take different forms. For intrinsic contribution, $\mathcal{B}(\omega,T)$ is the Berry curvature density given by electronic band structures or the Fermi surface topology \cite{Haldane2004PRL,Xiao2010RMP}. For extrinsic skew scattering, it is related to the magnetization and the scattering rate \cite{Fert1987,Nagaosa2010RMP}. Because of these complications, the behavior of transverse transport is largely unknown except in very special cases. For example, at zero temperature, it has been argued that the quasiparticle Nernst coefficient is related to the carrier mobility ($\mu$) and the Fermi energy ($\epsilon_F$) via $\nu/T=\pi^2\mu/3\epsilon_F$. This simple relation seems to hold over six orders of magnitude for a large spectrum of quantum materials \cite{Behnia2009}, but at finite temperatures, no such universal relation is known, and numerous puzzling experimental data have been accumulated that are too anomalous to interpret.

The mystery lies in the fact that most studies assume a temperature much smaller than the carrier bandwidth, so that the transport properties rely heavily on microscopic details such as the Berry curvatures, the Fermi surface topology, or the scattering anisotropy. Consequently, their behaviors are highly non-universal depending on the variation of $\mathcal{B}(\omega,T)$  with energy. It is therefore quite unexpected that universality may actually emerge when the temperature reaches the bandwidth of the effective carriers that dominate the transverse transport so that microscopic details are thermally averaged out. Similar scaling law has been observed in the thermal Hall conductivity \cite{Yang2020PRL} and the anomalous Hall coefficient \cite{Yang2023arXiv} attributed to nonzero Berry curvatures in a narrow energy window associated with the pseudogap, which is much smaller than the  bandwidth of noninteracting electrons and therefore highly nontrivial. We develop below a unified scheme to analyze these transverse transport properties including the Nernst coefficient. 

To see how the scaling occurs, we assume a regular $\mathcal{B}(\omega,T)$ being nonzero only for $\omega\in[-D_h,D_h]$, where $D_h$ is an effective bandwidth of transverse charge carriers induced by the pseudogap. Applying the Taylor expansion and defining
\begin{equation}
L_n(t)=\frac{t^n}{n!}\int^{1/t}_{-1/t} dx\frac{x^n}{\cosh^2(x/2)},
\end{equation}
where $x=\omega/D_h$ and $t=T/D_h$, we rewrite the conductivity formula as
\begin{equation}
\frac{\sigma^\alpha_{xy}}{H}=\frac{\pi^2}{3t^{\alpha}}\sum^\infty_{n=0} b_n(t) L_{n+\alpha}(t)
\end{equation}
where $b_n(t)=\mathcal{B}^{(n)}(0,T)$ is the $n$-th derivative of $\mathcal{B}(\omega,T)$ assumed to be regular for all $n$. 

\begin{figure}[t]
\centering\includegraphics[width=0.5\textwidth]{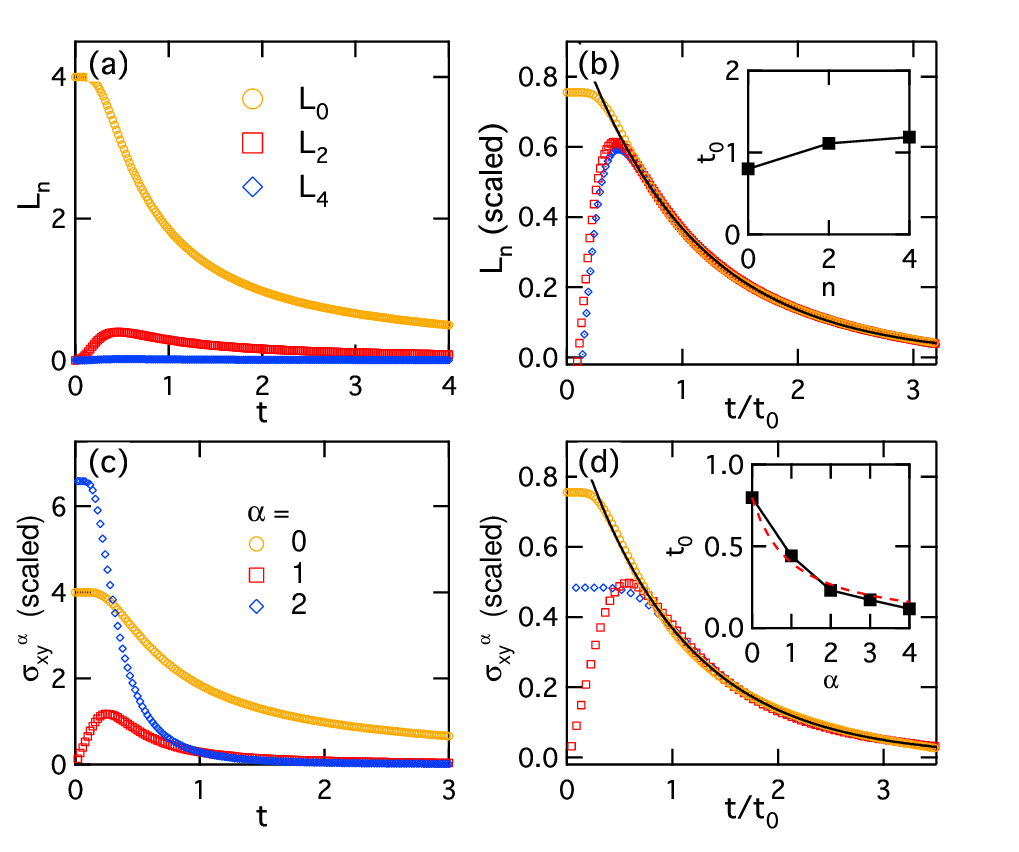}
\caption {Temperature dependence of (a) $L_n$ for $n=0$, 2, 4, and (c) $\sigma^\alpha_{xy}$ for $\alpha=0$, 1, 2 with constant $b_n$. (b) and (d) show their collapse after proper rescaling, where the solid lines are the exponential function, $e^{-t/t_0}$. The insets show how $t_0$ varies with $n$ or $\alpha$. The dashed line gives $t_0(\alpha=0)/(1+\alpha)$.}
\label{fig2}
\end{figure}

The integral $L_n$ are evaluated numerically and plotted in Fig. \ref{fig2}(a) for $n=0$, 2, 4. The odd terms are zero. $L_0$ saturates at low temperature and decreases continuously with increasing $t$, while all others approach zero as $t\rightarrow 0$ or $\infty$, and exhibit a maximum in between. These are expected since $L_n(t)\propto t^n$ for $t\rightarrow 0$ and $\propto t^{-1}$ for $t\rightarrow \infty$. In between, they look quite different and their magnitude decreases rapidly with increasing $n$. But surprisingly, as shown in Fig. \ref{fig2}(b), they can all be scaled to a single curve, $L_n\sim e^{-t/t_0}$, over the wide intermediate temperature region. The extracted $t_0=T_0/D_h$ (inset) is around unity and varies slightly with $n$.

The very different magnitude of $L_n$ implies that the major contribution to $\sigma^\alpha_{xy}$ comes from a single term in the expansion for regular choices of $b_n$. We have thus
\begin{equation}
\sigma^0_{xy}\propto b_0L_0,\ \ \ \ \sigma^1_{xy}\propto b_1\frac{L_2}{t},\ \ \ \ \sigma^2_{xy}\propto b_0\frac{L_2}{t^2}.
\end{equation}
To see how they behave, we first set $b_n$ to be independent of temperature. The results are plotted in Fig. \ref{fig2}(c). For $t\rightarrow0$, $\sigma^0_{xy}$ and $\sigma^2_{xy}$ saturate, while $\sigma^1_{xy}\propto t$. For $t\rightarrow\infty$, $\sigma^{\alpha}_{xy}\propto t^{-1-\alpha}$ for all $\alpha$. Strikingly, as plotted in Fig. \ref{fig2}(d), all curves again collapse onto the exponential function at intermediate temperatures, reflecting a special property of the basic formula. The scaling is not affected after divided by $t^\alpha$ except that the functions decay more rapidly at high temperatures to yield a smaller $t_0$. We see in the inset of Fig. \ref{fig2}(d) that $t_0$ is about $0.8$ for $L_0$ and $0.12$ for $L_4/t^4$, reduced by a factor of 7. Actually, expanding $t^{-\alpha}$ around $t_0$, we obtain $e^{-t/t_0}/t^{\alpha}\rightarrow e^{-t/t_0-\alpha t/t_0}$, which immediately yields $t_0\rightarrow t_0/(1+\alpha)$ and explains the reduction (dashed line).

We now turn to the Nernst coefficient \cite{Behnia2009}:
\begin{equation}
\nu=\frac{1}{H}\left(\frac{\sigma^1_{xy}}{\sigma^0_{xx}}-\frac{\sigma^1_{xx}}{\sigma^0_{xx}}\frac{\sigma^0_{xy}}{\sigma^0_{xx}}\right),
\end{equation}
which involves not only transverse quantities, but also the longitudinal resistivity $\rho=1/\sigma^0_{xx}$ and the Seebeck coefficient $S=\sigma^1_{xx}/\sigma^0_{xx}$. Both $\rho$ and $S$ depend on microscopic details of quasiparticle scattering and exhibit non-universal behavior in the intermediate temperature region. To simplify the discussion, we ignore the Seebeck term, which is relatively small in many cases, and consider some limits by taking $\rho\propto T^\beta$, where $\beta=0$ accounts for disorder, 1 for strange metal, and 2 for the Fermi liquid. We have then
\begin{equation}
\frac{\nu}{T}\propto b_1\frac{L_2}{t^{2-\beta}}.
\end{equation}
For constant $b_1$, all three situations ($\beta=0$, 1, 2) have already been shown in Fig. \ref{fig2}. For $\beta=0$, $\nu/T\propto L_2/t^2$ approaches a constant at zero temperature, while the other two curves, $L_2/t$ and $L_2$, are both suppressed as $t\rightarrow0$ and have a peak at around $t_0/2$. Regardless of these details, their collapse onto the same exponential function proves the robustness of the observed scaling in Fig. \ref{fig1}.

\begin{figure}[t]
\centering\includegraphics[width=0.45\textwidth]{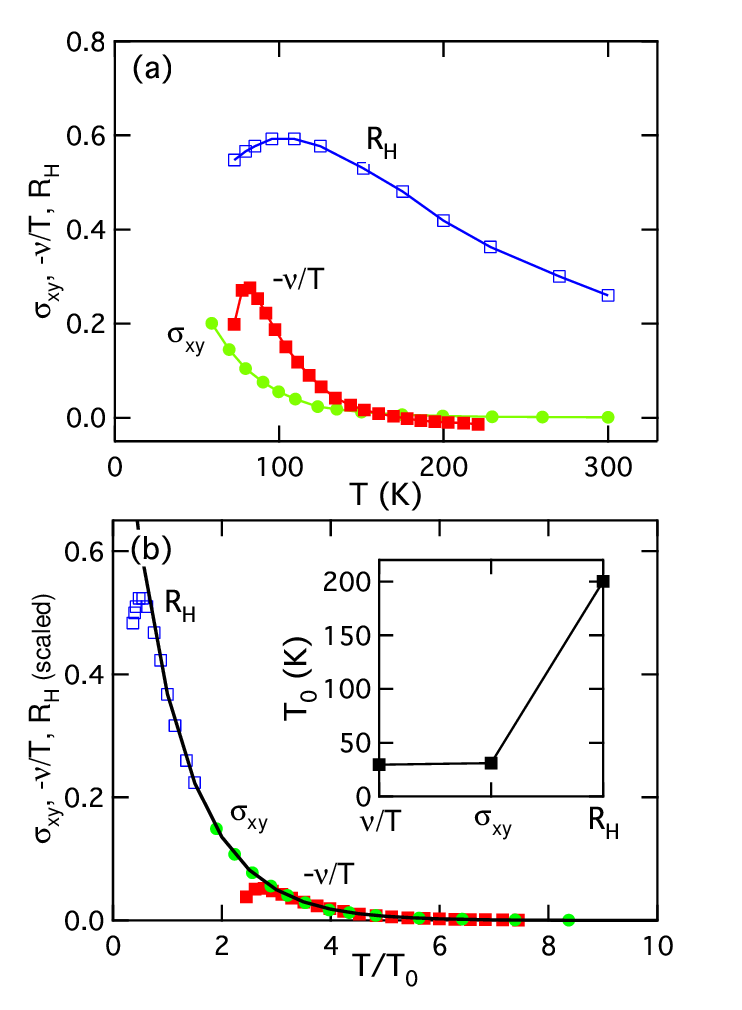}
\caption {(a) Comparison of the Nernst coefficient $\nu/T$ (nV K$^{-2}$ T$^{-1}$), the Hall conductivity $\sigma_{xy}$ (100 $\Omega^{-1}$ cm$^{-1}$), and the Hall coefficient $R_H$ (10$^{-2}$ cm$^{3}$ C$^{-1}$) for YBa$_2$Cu$_3$O$_y$ at a similar doping. The data are reproduced from experiment for the Hall conductivity at $y=6.60$ and the Hall coefficient at $y=6.65$ \cite{Segawa2004PRB}, and the Nernst coefficient along $a$-axis at $y=6.67$ \cite{Daou2010Nature}. (b) Collapse of all data on a single exponential function after rescaling. The solid line is the exponential function, and the inset shows the extracted values of $T_0$.}
\label{fig3}
\end{figure}

Including temperature dependence in $b_n$ won't destroy the scaling, but may have important consequences on the understanding of experimental measurements. This is best seen from the anomalous Hall coefficient, $R_H= \rho^2\sigma^{0}_{xy}\propto b_0t^{2\beta}L_0$, which diverges at large $t$ if $\beta>0.5$ and $b_0$ is independent of temperature. This contradicts with the experiments, where $R_H$ also obeys the exponential scaling and approaches a constant at high temperatures \cite{Hwang1994PRL,Yang2023arXiv}. Thus, $b_0$ must diminish as $t\rightarrow\infty$, which is conceivable since the quasiparticles are strongly correlated in underdoped cuprates and have a finite lifetime $\tau$ that decreases rapidly with temperature. In the semiclassical approximation \cite{Ong1991PRB,Narikiyo2020JPSJ}, the Boltzmann transport equation requires $\rho\propto \tau^{-1}\propto t^{\beta}$ and $\mathcal{B}\propto \tau^2$. Assuming $b_n\propto \tau^2\propto t^{-2\beta}$ leads to
\begin{equation}
R_H\propto L_0,\ \ \ \ \sigma^0_{xy}\propto \frac{L_0}{t^{2\beta}},\ \ \ \ \nu/T\propto \frac{L_2}{t^{2+\beta}}.
\end{equation}
Note that for $t\rightarrow 0$, disorder often dominates to give $\beta=0$ and recover the well-known $\nu\propto T$. Below we focus on the the exponential scaling at intermediate temperatures. Because $t_0$ is roughly unity for $L_n$, the above formulas predict that the exacted $T_0$ from $R_H$ gives roughly the effective quasiparticle bandwidth with nonzero Berry curvatures. For $\beta=2$ as in YBCO \cite{Segawa2004PRB}, $\sigma^0_{xy}$ and $\nu/T$ should have similar but much smaller $T_0$ since they are suppressed  by the same prefactor $t^{-4}$.

To see if these might be the case, we compare in Fig. \ref{fig3}(a) the experimental data of $\nu/T$, $\sigma_{xy}$, and $R_H$ for YBCO at a similar doping \cite{Daou2010Nature,Segawa2004PRB}. At first glance, they behave quite differently, with the Hall coefficient persisting to much higher temperature than others. However, as shown in Fig. \ref{fig3}(b), they all fall nicely onto the exponential scaling function. The extracted $T_0$ is given in the inset. We see similar values of roughly 30 K for $\nu/T$ and $\sigma_{xy}$, but  about 200 K for $R_H$. The latter is close to the onset temperature of the pseudogap from the resistivity \cite{Daou2010Nature}. It is quite amazing that the ratios between these numbers agree even quantitatively well with our simple scaling analysis. 

Additionally, we may apply the same analysis to the Hall angle and expect it to satisfy
\begin{equation}
\cot\Theta_H=\sigma^0_{xx}/\sigma^0_{xy}\propto \frac{t^\beta}{L_0},
\end{equation}
which predicts a crossover from $\cot\Theta_H\propto t^\beta$ at low temperatures to $\tan\Theta_H\propto e^{-t/t_0}$ at higher temperatures. Systematic analyses of experimental data have confirmed this crossover in YBCO \cite{Wuyts1996PRB,Yang2023arXiv}.

Disparity in the pseudogap temperature from different measurements has been a long-standing puzzle widely observed in cuprate experiments, causing a so-called large pseudogap from the susceptibility or the Hall coefficient and a small one from other transport measurements \cite{Timusk1999RPP,Luo2008PRB}. Our analysis suggests that their difference is nothing but a simple consequence of the temperature dependent prefactors. The excellent agreement of our theory with experiment confirms the consistency of our unified interpretation of transverse transport properties in hole-doped cuprates. It is then important to distinguish the true pseudogap transition temperature and the ill-defined ``onset" temperature of these observables. From our point of view, the latter is not a good indicator of the onset of the pseudogap phase and could be sometimes even misleading. 

Our idea of exploring universal scaling properties from the formula's structure may also have important implications in other strongly correlated or narrow band systems. In heavy fermion materials, a universal scaling indeed has been found for the Nernst coefficient, which follows closely the temperature evolution of  emergent heavy quasiparticles \cite{Yang2016RPP,Yang2020PRR}. Because the temperature is lower than the coherence temperature characterizing the heavy electron bandwidth, the above analysis might not apply. But it is wondering if this different scaling may actually fall into another interesting scenario where $\mathcal{B}(\omega,T)$ exhibits $\omega/T$ scaling, namely, $\mathcal{B}(\omega,T)=b(T)\mathcal{B}(\omega/T)$, possibly due to quantum criticality. We would then have
\begin{equation}
\frac{\sigma^\alpha_{xy}}{H}\approx \frac{\pi^2}{3}b(T)\int dx\frac{x^\alpha}{\cosh^2(x/2)}\mathcal{B}(x),
\end{equation}
such that $\sigma^\alpha_{xy}\propto b(T)$. More investigations are needed to elaborate on this possibility. In any case, it will be interesting to explore more potential scaling laws in the intermediate temperature region based solely on the mathematical structure of theoretical formulas instead of resorting to microscopic details, which might provide important information on some unexpected but quite generic properties of the underlying quasiparticles.

This work was supported by the National Key R\&D Program of China (Grant No. 2022YFA1402203), the National Natural Science Foundation of China (Grants No. 12174429, No. 11974397),  and the Strategic Priority Research Program of the Chinese Academy of Sciences (Grant No. XDB33010100).

\clearpage

\end{document}